\documentstyle[epsfig,longtable]{aipproc}
\def\ga{\;\rlap{\lower 2.5pt
 \hbox{$\sim$}}\raise 1.5pt\hbox{$>$}\;}
\def\la{\;\rlap{\lower 2.5pt
   \hbox{$\sim$}}\raise 1.5pt\hbox{$<$}\;}
\def\gsim{\;\rlap{\lower 2.5pt
 \hbox{$\sim$}}\raise 1.5pt\hbox{$>$}\;}
\def\lsim{\;\rlap{\lower 2.5pt
   \hbox{$\sim$}}\raise 1.5pt\hbox{$<$}\;}

\begin{document}
\begin{flushright}
{\footnotesize
FERMILAB-Conf-98/378-A}
\end{flushright}
\nopagebreak
\vspace{-\baselineskip}
\title{Models for High--Redshift Ly$\alpha$ Emitters}
 
\author{Z. Haiman$^*$ and M. Spaans$^{\dagger}$}
\address{$^*$Astrophysics Theory Group, Fermi 
National Accelerator Laboratory, Batavia, IL 60510\\
$^{\dagger}$Harvard Smithsonian Center for Astrophysics,
60 Garden Street, Cambridge, MA 02138}

\lefthead{High Redshift Lyman $\alpha$ Emitters}
\righthead{Haiman \& Spaans}
\maketitle

\begin{abstract}
We present models for dusty high--redshift Ly$\alpha$ emitting galaxies by
combining the Press--Schechter formalism with a treatment of inhomogeneous dust
distribution inside galaxies.  These models reproduce the surface density of
emitters inferred from recent observations, and also agree with previous
non--detections.  Although a detailed determination of the individual model
parameters is precluded by uncertainties, we find that (i) the dust content of
primordial galaxies builds up in no more than $\sim 5\times 10^8$ yr, (ii) the
galactic HII regions are inhomogeneous with a cloud covering factor of order
unity, and (iii) the overall star formation efficiency is at least $\sim5\%$.
Future observations should be able to detect Ly$\alpha$ galaxies upto redshifts
of $z\sim8$. If the universe is reionized at $z_{\rm r}\lsim 8$, the
corresponding decline in the number of Ly$\alpha$ emitters at $z\gsim z_{\rm
r}$ could prove to be a useful probe of the reionization epoch.
\end{abstract}

\section*{Introduction}

Until a decade ago, the search for high--redshift Ly$\alpha$ galaxies had
enjoyed no compelling successes~\cite{tdt95}.  With the improved sensitivity on
large area telescopes in the last couple of years, these young galaxies are
finally being detected~\cite{hcm98,d98}.  Clearly, this population is of great
interest to the field of galaxy formation and the early evolution of the
universe.  Fundamental questions we need to understand are why the earlier
surveys have been unsuccessful, how many objects are still expected to be
found, and what physical conditions pertain in these early systems so that the
Ly$\alpha$ radiation may escape.

In a galactic setting, the emerging luminosity of the Ly$\alpha$ line is
strongly modulated by the amount and spatial distribution of stellar
dust~\cite{n91}. Observations have firmly established the strong decrease in
Ly$\alpha$ equivalent width with increasing oxygen abundance~\cite{t93}.
Indeed, even a modest amount of dust (observationally traced by oxygen) inside
a homogeneous HII region is sufficient to attenuate all of the produced
Ly$\alpha$ radiation, because Ly alpha photons are resonantly scattered, 
and dust absorption significantly increases the effective line optical
depths~\cite{cf93}.  However, this situation can be alleviated if the medium is
inhomogeneous~\cite{n91}.  In a multi--phase medium where dust resides in
opaque clumps, the photons do not penetrate the clumps and spend most of their
time in the interclump medium, where the opacity is very small.  The result is
that the escape of Ly$\alpha$ radiation is significantly enhanced compared to
the case of resonant scattering in a homogeneous medium.  Alternatively, large
enough velocity gradients can provide a second way of efficient escape of
Ly$\alpha$ radiation, as suggested by recent observations of 4 nearby
galaxies~\cite{k98}.
 
In a cosmological context, the formation of dust requires the presence of
metals.  Since the enrichment of the gas with metals is expected to build up
only gradually during the star formation history of the universe~\cite{hl97},
at high enough redshifts the enrichment level will be low.  The correspondingly
small dust content of the earliest galaxies would facilitate the escape of
their Ly$\alpha$ radiation.  Such an interplay between the dust enrichment of
the IGM and the galactic environment could be the mechanism that leads to the
observed surface density of high--z Ly$\alpha$ emitters~\cite{hs98}.

\section*{Ly$\alpha$ Emission from Inhomogeneous Clouds}

\begin{figure} 
\centerline{\epsfig{file=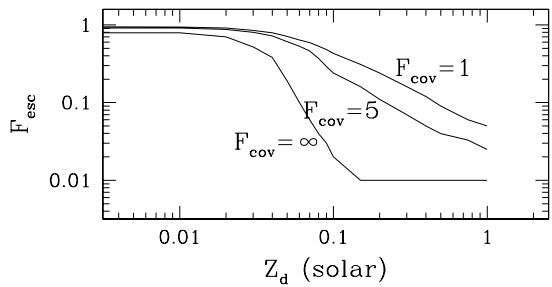,width=3.in,height=3.in}}
\vspace{-1.5in}
\vspace{10pt}
\caption{ {\it The escape fraction $F_{\rm esc}$ of Ly$\alpha$ photons from an
inhomogeneous medium, as a function of dust content.  The three curves
correspond to three different values of the covering factor of opaque clumps,
$F_{\rm cov}=$1, 5, and $\infty$.}  }\label{fig:fesc}
\end{figure}

The radiative transfer problem which needs to be solved for the Ly$\alpha$ line
is well studied~\cite{n90}.  We have modeled individual galaxies with a range
of masses for the ionizing stars, dust content, and inhomogeneity, using a
numerical Monte Carlo approach~\cite{s96}.  We assumed a Scalo IMF for the
spectral types of the central stars in the HII regions, ranging from O5 to B1,
with the stars distributed in a statistically homogeneous manner inside a
percolating multi--phase medium.  We adopted the formalism of
Neufeld~\cite{n91} for a multi--phase medium, and assumed a typical line width
of $\Delta v \sim8$ km/s.  We parameterized the multi--phase medium by opaque
clumps embedded in an inter--clump medium of negligible opacity.  The clump
covering factor $F_{\rm cov}$, i.e.\ the average number of clumps along a line
of sight, fixes the degree of inhomogeneity.  The escape fractions $F_{\rm
esc}$ were computed on a grid of models with dust contents between $Z_{\rm
d}=10^{-2}-1$ solar, and covering factors $F_{\rm cov}=1,5,\infty$.
Figure~\ref{fig:fesc} shows the escape fraction, and demonstrates its
sensitivity to the covering factor. The inhomogeneous percolating slabs are
much more transparent than the homogeneous ones; the difference around $Z_{\rm
d}\sim10^{-1}$ solar is over an order of magnitude.

\section*{Cosmological Abundance of Ly$\alpha$ Emitters}

\begin{figure} 
\centerline{\epsfig{file=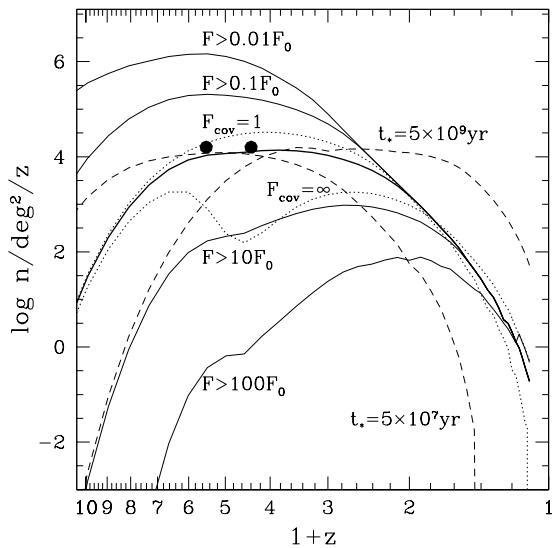,width=3.in,height=3.in}}
\vspace{10pt}
\caption{{\it The surface density of Ly$\alpha$ emitters in our standard model
(solid lines) with fluxes above different values of the detection threshold.
The two data points are taken from~\protect\cite{hcm98}. For the fixed
threshold $F_0=1.5\times10^{-17}~{\rm erg~cm^{-1}~s^{-1}}$, the dashed lines
show the surface density when the star--formation rate is increased or
decreased by a factor of 10.  Similarly, the dotted lines show the surface
density when the covering factor is changed to $F_{\rm cov}=1$, or $\infty$.}}
\label{fig:res1}
\end{figure}

In order to model the cosmological abundance of high--redshift Ly$\alpha$
emitters, we assumed that the formation of dark matter halos follows the
Press--Schechter~\cite{ps74} theory, and every halo with a virial temperature
larger than $10^4$ K forms a galaxy that goes through a Ly$\alpha$ emitting
phase~\cite{hrl97}.  A fraction $\epsilon_\star$ of the gas was turned into
stars, with a constant star formation rate (SFR) over a period of $t_\star$
years.  The SFR was related to the intrinsic Ly$\alpha$ luminosity~\cite{k93},
assuming case B recombination.  We adopted the simplest assumption, i.e.\ that
$\epsilon_\star$ and $t_{\star}$ both have the same constant values in each
halo. The amount of dust produced and retained in each galaxy, $Z_{\rm d, ISM}$
then increases linearly for $t_\star$ years, after which it reaches the final
value of $Z_{\rm d,ISM}(t_\star)=0.3$ solar.  Similarly, we assume that each
galaxy deposits dust into the surrounding IGM at a constant rate for $t_\star$
years, and causes an enrichment of the {\it surrounding regions} within the
intergalactic medium to $Z_{\rm d, IGM}(z)$.  We normalize $Z_{\rm d, IGM}$ to
the cluster metallicity of $\sim 0.3$ solar at redshift $z\sim 1$. Note that
our $Z_{\rm d, IGM}$ is the average metallicity within the polluted regions,
and not the universal average dust fraction of the IGM. Given a cosmology, the
five parameters $t_\star$, $F_{\rm cov}$, $\epsilon_\star$, $Z_{\rm d,
IGM}(z=1)$, and $Z_{\rm d, ISM}(t_\star)$ uniquely determine the number density
of Ly$\alpha$ emitters at any flux and redshift.  Figure~\ref{fig:res1} shows
the resulting surface density of emitters, as a function of redshift, in a flat
$\Lambda$CDM model with a tilted power spectrum ($\Omega_0,\Omega_\Lambda,
\Omega_{\rm b},h,\sigma_{8h^{-1}},n$)=(0.35, 0.65, 0.04, 0.65, 0.87, 0.96).  We
also indicate the observed surface density of emitters~\cite{hcm98} at the two
redshifts $z=3.4$ and $z=4.5$.

Figure~\ref{fig:res1} has several interesting features.  It demonstrates that
the surface abundance strongly depends on the time over which the dust is
produced ($t_{\rm star}$), and on the ambient inhomogeneity of the HII regions
that surround the ionizing OB stars ($F_{\rm cov}$).  Our models indicate that
the dust content builds up in $\lsim 5\times 10^8$ yr, the galactic HII regions
are inhomogeneous with a cloud covering factor of order unity, and the overall
star formation efficiency is at least $\sim5\%$.  These numbers should be
predicted by more complete and detailed models of galactic evolution, and will
be useful discriminators between such models.  The surface density of emitters
is also a steep function of the detection threshold, a feature that makes our
model consistent both with recent detections~\cite{hcm98} and earlier upper
limits~\cite{tdt95}.  Our models predict that the surface density changes
relatively slowly with redshift to $z\lsim8$, and Ly$\alpha$ galaxies will be
detectable, around the present flux threshold, upto redshifts as high as
$\sim$8. Recent detections of three new Ly$\alpha$ emitters at $z\sim5.7$
support this conclusion~\cite{c98}.  However, if the universe is reionized at
$z_{\rm reion}\lsim8$, then the damping wing of Ly$\alpha$ absorption from the
neutral IGM would severely damp the Ly$\alpha$ emission line~\cite{m98}. This
would render Ly$\alpha$ emission from $z\gsim z_{\rm reion}$ undetectable, and
cause a decline in the number of observed emitters beyond $z_{\rm reion}$.  In
addition, the neutral IGM would likely imprint a characteristic assymetry on
the emission line profiles of emitters located near the reionization epoch.
Provided this assymetry is measured, the disappearance of the Ly$\alpha$
emitter population could be useful diagnostic signature of reionization.

ZH was supported at Fermilab by the DOE and the NASA grant NAG 5-7092.

\end{document}